# Development of a mini-PET Detector based on Silicon Photomultiplier Arrays for Plant Imaging Applications

F. Barbosa, H. Dong, B. Kross, S.J. Lee, Y. Mack, J. McKisson,
J. McKisson, A Weisenberger, W. Xi, C. Zorn
Thomas Jefferson National Accelerator Laboratory, Newport News, VA 23606

S. Majewsk, A. Stolin
West Virginia University, Morgantown, WV 26505

C.R. Howell, A.S. Crowell, C.D. Reis
Duke University, Durham, NC 27708

M.F. Smith
University of Maryland, Baltimore, MD 21201

*Abstract* - A mini-PET style detector system is being developed for a plant imaging application with a compact array of silicon photomultipliers (SiPM) replacing position sensitive photomultipliers (PSPMT). In addition to compactness, the use of SiPMs will allow imaging setups involving high strength MRI-type magnetic fields. The latter will allow for better position resolution of the initial positron annihilations in the plant tissue. In the present work, prototype arrays are tested for the uniformity of their response as it is known that PSPMTs require significant gain compensation on the individual channels to achieve an improved uniformity in response. The initial tests indicate a high likelihood that the SiPM arrays can be used without any gain compensation.

## INTRODUCTION

The use of radiotracers to track various biological activities in plants has been a steadily developing field. One important component has been the interdisciplinary modification of medical bioimaging detector systems for the specific needs of plant biology research. Reviews of some of this work can found in the literature [1], [2]. As a specific example, a collaboration at Duke University involving the Triangle Universities Nuclear Laboratory (TUNL) and the Duke University Phytotron has been using $^{11}CO_2$ tracers to study the dynamical responses of plants to environmental changes to elevated CO2 levels in the atmosphere. Other groups in this area include Brookhaven National Laboratory (BNL) in Long Island, New York, Forschungszentrum Jülich (Germany), and the Japan Atomic Energy Agency (Japan). So far, only the Duke/TUNL group has built a system uniquely adapted to the needs of the plant biology research as opposed to modifying an existing biomedical imaging system [2]. The work has indicated the need to develop unique imaging technologies similar to those used in biomedical work (such as PET) but with characteristics developed for the unique and particular character of plant biology research. In a previous paper [3], an initial detector design by the JLAB group has been tested at the TUNL/Phytotron facility. Based upon this work, a new set of detector designs is being developed based on a planar PET design [4].

## PRESENT PMT-BASED DESIGN

The chief radiotracer used in the TUNL/Phytotron studies is $^{11}CO_2$ where the $^{11}C$ has a half-life of about 20 minutes. The emitted positron has a mean energy of 326 keV and maximum of 959 keV. In water, the range is 1.1 mm on average with a maximum range of 4.1 mm [1]. A PET design would detect the two annihilation photons formed after the positron achieves thermal energies. One of the requirements is the need to be able to modify the PET setup according to the specifics of the plant setup. Fig. 1 shows a new "lego" design where individual 5x5 cm$^2$ detector modules (Hamamatsu H8500 PSPMTs) could be stacked into a variable size array that can also be arranged either in a planar form or a circular arc. Each PSPMT would be coupled to a LYSO:Ce scintillating crystal array (1x1x6 mm$^3$). Fig. 2 shows photographs of the actual modules in two configurations. This system will be tested in the near future.

One ongoing problem with using the positron radiotracer is that given the relative thinness of the plant tissues, the emitted positron has a high probability of traveling a significant distance before annihilating into two photons. One possible way to deal with this would be to use a high strength magnetic field in the multi-Tesla range to force the positron to spiral in a small enclosed path before annihilation, thereby restricting its linear range from the point of initial emission. This would require the use of magnetically immune photodetectors that also have the superior photodetection capabilities of vacuum photomultipliers. This now seems possible with the ongoing development of the devices known colloquially as "silicon photomultipliers" (SiPM) or more technically as limited Geiger mode avalanche photodiodes [5].

At present the project is focused upon the development of a flexible array using Hamamatsu H8500 PSPMTs coupled to LSYO:Ce scintillator arrays (Fig. 1). With this "lego"-like design, it will be possible to vary both the size of the detector array as well as to arrange the detectors in a optimum configuration other than simply a planar setup. The SiPM version will initially be a simple pair of planar detectors used in a PET mode. A design similar to the PSPMT version will also be investigated.

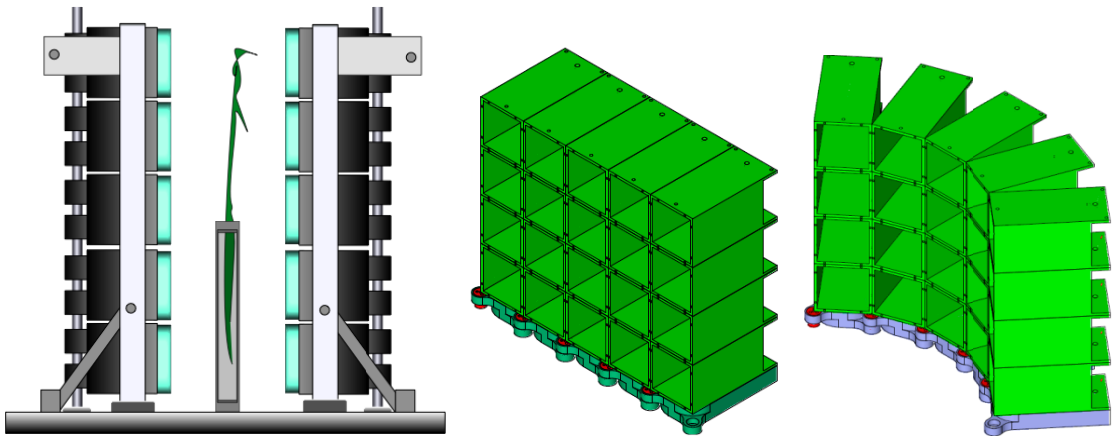

Fig. 1 (Left) Conceptual drawing of dual planar detector for H8500/LYSO setup. (Middle and Right): Sketches of the modules (with no detectors) in planar and circular setup.

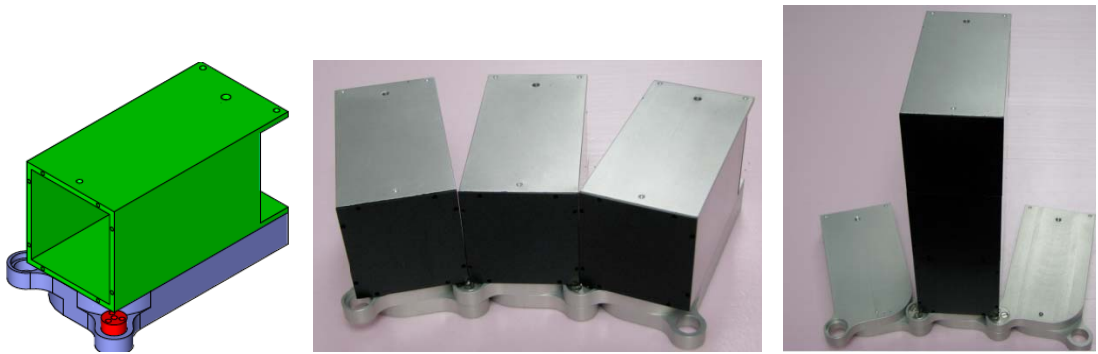

Fig. 2: Photographs of modules next to a schematic of same at left.

## INITIAL TESTS OF THE SIPM ARRAYS

In this paper, the focus will be on the initial tests of two samples of the SiPM arrays as shown in Fig. 3. The first question that has arisen has been the analogue of the initial problems when PSPMTs were tested. That is, the inter-channel differences in gain across the surface of the PSPMT could be considerable, even reaching differences of 3-to-1 in some cases. The problem was resolved by creating a readout circuit that used an array of precision resistors to compensate for the differences among the channels of a single PSPMT. An analogous concern was also raised with SiPM arrays. In actual fact, the vendor provided a complete list of the recommended operating voltages (to produce an identical gain of $7.5 \times 10^5$) for each of the 16 3x3 mm$^2$ units in all of the 20 arrays. The actual variation was quite small (a few tens of volts), but it was decided to test the concept of a compensated array versus one that was used as provided.

Fig. 4 shows a schematic of the Hamamatsu array readout. The implemented readout circuit used a Center-of-Gravity (COG) mode with the rows/columns (X/Y) of the array split and summed. An amplification of x4 was also provided for each of the 16 elements. For one of samples (module 3), an additional set of precision resistors was added to the readout board to compensate for the difference in gain among the 16 elements since the array was operated at a single bias voltage. The other sample (module 4) was left unmodified.

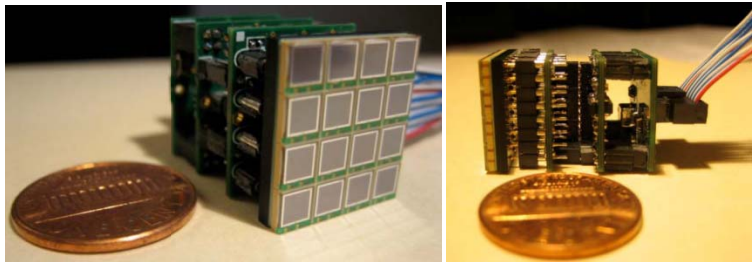

Fig. 3: Photographs of the Hamamatsu S11064-050P array attached to the in-house (JLAB) readout board. A COG method is implemented in the readout for this version. A more efficient resistive readout system will be implemented in the full planar detector.

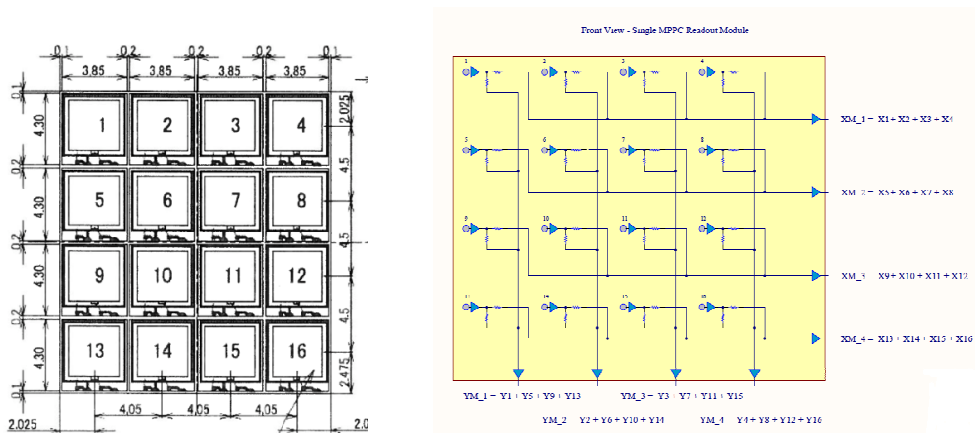

Fig. 4: (Left) Schematic of the S11064-050P array. It uses 16 3x3 mm$^2$ elements of the S10931-050P. These have 3600 pixels of 50 μm size. PDE at the recommended bias is in the range of 25-30% for wavelengths in the range of 450-470 nm. (Right) Schematic of the cog readout for the array. An additional amplification of x4 is also provided for each element of the array.

The first test consisted of illuminating the arrays with a very uniform source of fast, pulsed light (Fig. 5). This was accomplished with a blue LED [6] excited by a 10 ns width pulse from a pulse generator. This provided a fast pulse only slightly slower than one from a standard plastic scintillator. Since this light is highly directional, it was passed through a holographic diffuser [7] and a standard integrating sphere to produce a broadly spread, uniform illumination of the SiPM surface. From other studies of 1x1 mm$^2$ and 3x3 mm$^2$ SiPMs of the same type, the light intensity was set well within the linear response region of the photodetector array. Fig. 6 shows the results from illuminating the two array samples. The compensated array (module 3) shows a clear difference from the non-compensated module (#4). The largest differences are seen in the "Y" columns of the readout. This can traced back to the fact that the manufacturer place the individual elements in increasing order of optimum bias voltage, thereby creating an ordered difference in gain.

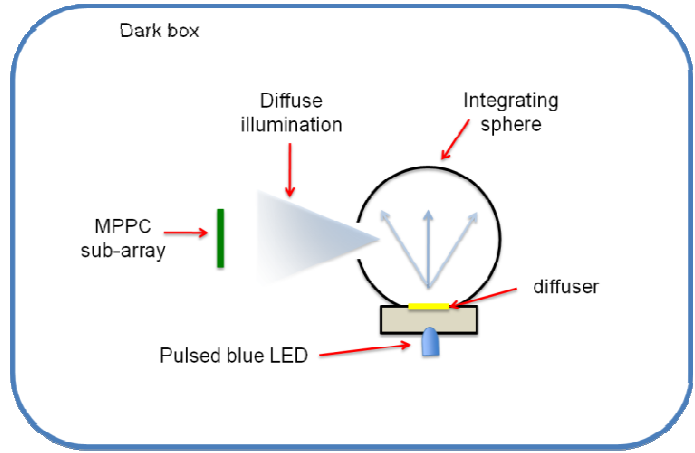

Fig. 5: Schematic of setup to test SiPM array uniformity of response. Pulsed LED has light put through diffuser and integrating sphere to produce a diffuse and uniform illumination of array. Integrating sphere also decreases intensity to allow SiPM to operate well within linear range of response.

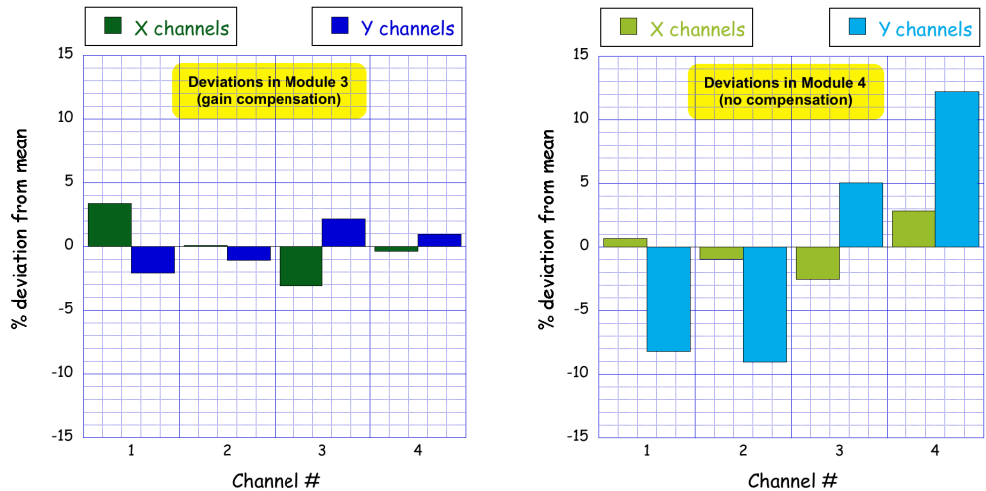

Fig. 6: Results of uniformity tests. (Left) Module 3 has array of compensating resistors to reduce differences in gain among the elements when operated at a single bias voltage. (Right) Module 4 was left without any compensation. The results indicate the possible need of compensating resistors as with the H8500 pmts.

The next step was to use an actual scintillating crystal to test the ability of the arrays to discern the individual pixels of the array. Fig. 7 shows a schematic of the setup. Past experience has shown that it is necessary to use a specific thickness of glass spacer to minimize the pincushion-like distortion seen when the light from the scintillator pixels is not properly shared among the elements of the photodetector. An optimum thickness of 3.2 mm created a uniform image. A thicker spacer maintained the uniformity, but decreased the resolution of the pixel elements. It was known from other work that a 1x1 mm$^2$ pixel size would provide a severe challenge, so it was decided to test the arrays with a 1x1x5 mm$^3$ LYSO crystal only slightly larger than the array size. Fig. 8 and 9 shows the results for both arrays. It can be seen that both images look very similar regardless of the use of gain compensation. To observe the more subtle differences, a profile slice across each of images is also shown. Again, the differences are minimal. As a comparison, Fig. 10 shows a equivalent result from using a H8500 PSPMT with a scintillating array. The H8500 used a compensating circuit to remove the gain differences among the 64 channels. Despite the great improvement with this technique, there is still residual change in uniformity across the image that is not seen with the SiPM arrays.

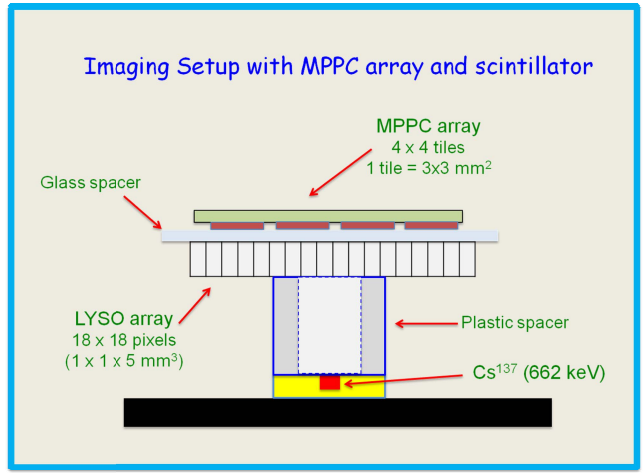

Fig. 7: Setup used to test imaging capability of the two arrays. A gamma source illuminates a finely pixilated LYSO array (1x1x5 mm³). The glass spacer allows for proper light sharing among the array elements and reduces the distortion in the final image.

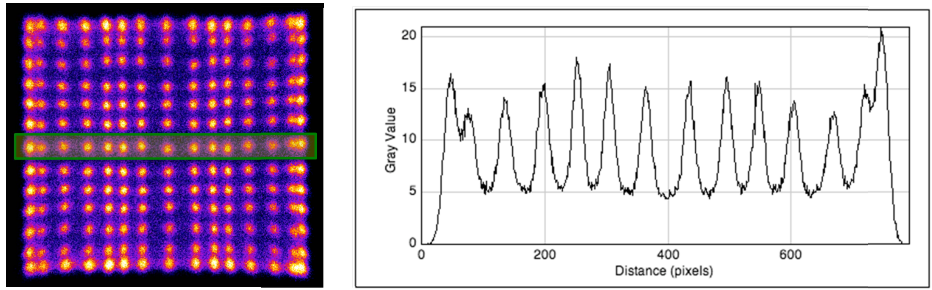

Fig. 8: Results of imaging test for compensated module 3. Left is the 2D image. Right is a 1D profile through the middle of image. The LYSO crystal is slightly larger than the SiPM array, so some pixels are not detected.

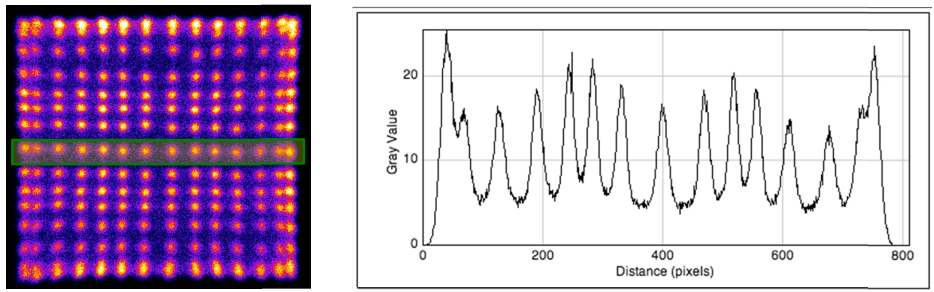

Fig. 9: The analogous result for the non-compensated module 4. Again, a uniform image is formed even without any gain compensation.

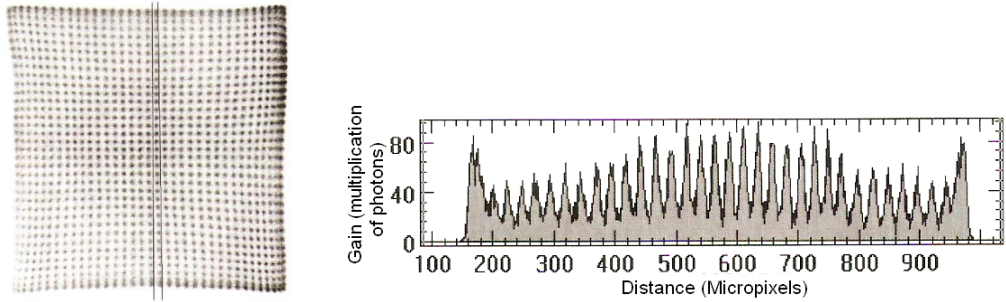

Fig. 10: A similar result taken from a previous study [8] using gain compensated H8500 phototubes. Even after compensation (for large differences), there is still a greater residual non-uniformity compared to the SiPM arrays.

With this promising result, the next step will be to create the two full size planar detectors, each using a 3x3 array of the individual sub-arrays. No compensation circuits will be used. However, the readout method will be changed. The COG scheme would result in 24 readout channels per planar detector. Instead a resistive scheme will be implemented, reducing the readout to 4 channels (+ 1 sum) per detector. In conjunction with the other plant imaging studies, these planar detectors will be used in the plant radiotracer imaging program. There is a special interest in using these detectors in a strong MRI-type field. As mentioned previously, the positrons have a significant probability in exiting the plant tissue before annihilating, so it will be important to attempt to restrict the pre-annihilation flight path of the positrons within the plant.